\documentclass[a4paper]{report}
\usepackage[utf8]{inputenc}
\usepackage[T1]{fontenc}
\usepackage{RJournal}
\usepackage{amsmath,amssymb,array}
\usepackage{booktabs}

\providecommand{\tightlist}{%
  \setlength{\itemsep}{0pt}\setlength{\parskip}{0pt}}

\begin{document}

\sectionhead{Contributed research article}
\volume{XX}
\volnumber{YY}
\year{20ZZ}
\month{AAAA}

\begin{article}
\title{autoplotly - Automatic Generation of Interactive Visualizations for
Popular Statistical Results}
\author{by Yuan Tang}

\maketitle

\abstract{%
The \pkg{autoplotly} package provides functionalities to automatically
generate interactive visualizations for many popular statistical results
supported by \pkg{ggfortify} package with \pkg{plotly} and \pkg{ggplot2}
style. The generated visualizations can also be easily extended using
\pkg{ggplot2} and \pkg{plotly} syntax while staying interactive.
}

\section{Background}

With the help of base graphics, grid graphics, and \CRANpkg{lattice}
graphics \citep{lattice}, R users already have many plotting options to
choose from. Each has their own unique customization and extensibility
options. Nowadays, \CRANpkg{ggplot2} has emerged as a popular choice for
creating visualizations \citep{wickham2009ggplot2} and provides a strong
programming model based on a ``grammar of graphics'' which enables
methodical production of virtually any kind of statistical chart. The
\pkg{ggplot2} package provides a suit of succinct syntax and independent
components and makes it possible to describe a wide range of graphics.
It's based on an object-oriented model that is modular and extensible,
which becomes a widely used framework for producing statistical graphics
in R.

The distinct syntax of \pkg{ggplot2} makes it a definite paradigm shift
from base and \pkg{lattice} graphics and presents a somewhat steep
learning curve for those used to existing R charting idioms. Many
industry R users, especially the users that build web applications in R
by leveraging \CRANpkg{shiny} \citep{shiny} package, may not be
satisfied with static plots. Those web applications often involve user
interactions so that users can dive into the plots, explore areas of
interest, and select relevant data points for more details.
\CRANpkg{ggiraph} \citep{ggiraph} is an extention of \pkg{ggplot2} that
provides building blocks for users to build interactive plots and when
used within a shiny application, elements associated with an id can be
selected and manipulated on client and server sides. There are also
other packages such as \CRANpkg{d3r} \citep{d3r} and \CRANpkg{plotly}
\citep{plotly} built on top of Javascript visualization frameworks that
are totally isolated from \pkg{ggplot2} but become popular building
blocks for creating interactive visualizations in R.

Often times users only want to quickly iterate the process of exploring
data, building statistical models, and visualizing the model results,
especially the models that focus on common tasks such as clustering and
time series analysis. Some of these packages provide default base
\code{plot} visualizations for the data and models they generate.
However, they look out-of-fashion and these components require
additional transformation and clean-up before using them in
\pkg{ggplot2} and each of those transformation steps must be replicated
by others when they wish to produce similar charts in their analyses.
Creating a central repository for common/popular transformations and
default plotting idioms would reduce the amount of effort needed by all
to create compelling, consistent and informative charts. The
\CRANpkg{ggfortify} \citep{rjggfortify} package provides a unified
\pkg{ggplot2} plotting interface to many statistics and machine-learning
packages and functions in order to help these users achieve
reproducibility goals with minimal effort. \pkg{ggfortify} package has a
very easy-to-use and uniform programming interface that enables users to
use one line of code to visualize statistical results of many popular R
packages using \pkg{ggplot2} as building blocks. This helps
statisticians, data scientists, and researchers avoid both repetitive
work and the need to identify the correct \pkg{ggplot2} syntax to
achieve what they need. Users are able to generate beautiful
visualizations of their statistical results produced by popular packages
with minimal effort.

The \CRANpkg{autoplotly} \citep{autoplotly} package is an extension
built on top of \pkg{ggplot2}, \pkg{plotly}, and \pkg{ggfortify} to
provide functionalities to automatically generate interactive
visualizations for many popular statistical results supported by
\pkg{ggfortify} package with \pkg{plotly} and \pkg{ggplot2} style. The
generated visualizations can also be easily extended using \pkg{ggplot2}
and \pkg{plotly} syntax while staying interactive.

\section{Software Architecture}

The \pkg{autoplotly} package calls \pkg{ggfortify}'s \code{autoplot()}
method that invokes an registered S3 generic functions
\footnote{\url{http://adv-r.had.co.nz/S3.html}} for the applied object
to create the visualizations with \pkg{pplot2} style. Next, the
generated \code{ggplot} object is translated to \code{plotly} object
with interactive graphical components leveraging
\code{plotly::ggplotly}. Additional clean-up and correction are then
performed due to the feature parity between \pkg{plotly} and
\pkg{ggplot2} that results in redundant and corrupted components . For
example, if we want to generate interactive visualization for principal
components analysis results produced from \code{prcomp(...)}, the
following will be executed in order:

\begin{itemize}
\tightlist
\item
  \code{autoplotly(prcomp(...))} - calls \pkg{autoplotly}'s main
  function
\item
  \code{autoplot.prcomp(prcomp(...))} - invokes the registered S3
  generic function
\item
  \code{ggplotly(autoplot.prcomp(prcomp(...)))} - translates
  \code{ggplot} object to \code{plotly} object
\end{itemize}

The final object is of class \code{plotly} with the corresponding
\code{ggplot} object as one of its attributes. It can be easily extended
using either \pkg{plotly} or \pkg{ggplot2} style. When additional
\pkg{ggplot2} elements or components are applied, for example:

\begin{Schunk}
\begin{Sinput}
p <- autoplotly(prcomp(iris[c(1, 2, 3, 4)]), data = iris,
  colour = 'Species', label = TRUE, label.size = 3, frame = TRUE)

p <- p +
  ggplot2::ggtitle("Principal Components Analysis") +
  ggplot2::labs(y = "Second Principal Components", x = "First Principal Components")
p
\end{Sinput}
\end{Schunk}

The above example adds title and axis labels to the originally generated
plot. When \code{`+` <- function(e1, e2)} operator is applied to a
\code{ggplot} element as the second argument \code{e2}, e.g.
\code{ggplot2::ggtitle(...)}, the \code{ggplot} object that we attached
to the output of \code{autoplotly()} earlier will be used as the first
argument \code{e1}, borrowing \pkg{ggplot2}'s extensibility.

\begin{figure}[htbp]
  \centering
  \includegraphics[width=145mm,scale=0.8]{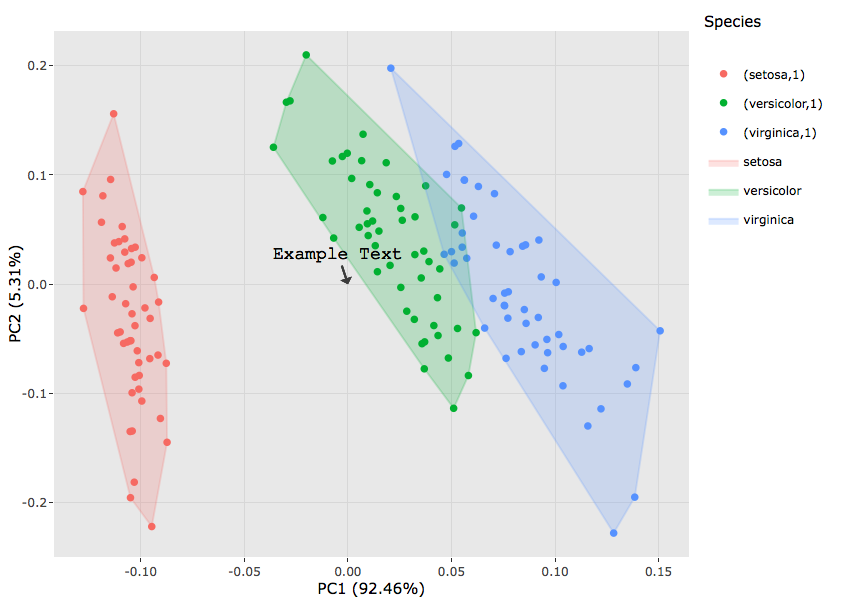}
  \caption{PCA with custom \pkg{plotly} style annotation element.}
  \label{figure:pca_plotly_annotation}
\end{figure}

Similarly, if we are adding \pkg{plotly} interactive components, the
\code{plotly} object from the output of \code{autoplotly()} \code{p}
will be used instead. The following code adds a custom \pkg{plotly}
annotation element placed to the center of the plot with an arrow, as
shown in Figure \ref{figure:pca_plotly_annotation}:

\begin{Schunk}
\begin{Sinput}
p <- autoplotly(prcomp(iris[c(1, 2, 3, 4)]), data = iris,
  colour = 'Species', frame = TRUE)

p %>% layout(annotations = list(
  text = "Example Text",
  font = list(
    family = "Courier New, monospace",
    size = 18,
    color = "black"),
  x = 0,
  y = 0,
  showarrow = TRUE))
\end{Sinput}
\end{Schunk}

\pkg{autoplotly} re-exports \code{plotly::subplot()} to enable users to
stack multiple interactive plots generated via \code{autoplotly()}
together. Some statistical results produce multiple plots, e.g.
\code{lm()} fitted model objects, which are given extra attention and
additional manipulations are performed in order to make sure users can
choose whether to share axis labels, change margins, and change layout
strategy while keeping the interactive control of multiple sub-plots
independently.

The snapshots of the interactive visualizations generated via
\code{autoplotly()} can be exported using a simple \code{export()}
function, e.g. \code{export(p, "/tmp/result.png")}. It builds on top of
\CRANpkg{RSelenium} \citep{RSelenium} for exporting WebGL plots and uses
\CRANpkg{webshot} \citep{webshot} for non-WebGL formats such as JPEG,
PNG, PDF, etc.

\section{Illustrations}

As demonstrated earlier, \pkg{autoplotly} package provides a
\code{autoplotly()} function to work with objects of different classes
produced from various popular statistical packages. This section
highlights some of the example of automatic interactive visualizations
from different types of statistical results.

\subsection{Principal component analysis}

The \code{autoplotly()} function works for the two essential classes of
objects for principal component analysis (PCA) obtained from \pkg{stats}
package: \code{stats::prcomp} and \code{stats::princomp}, for example:

\begin{Schunk}
\begin{Sinput}
autoplotly(prcomp(iris[c(1, 2, 3, 4)]), data = iris, frame = TRUE, colour = 'Species')
\end{Sinput}
\end{Schunk}

\begin{figure}[htbp]
  \centering
  \includegraphics[width=145mm,scale=0.8]{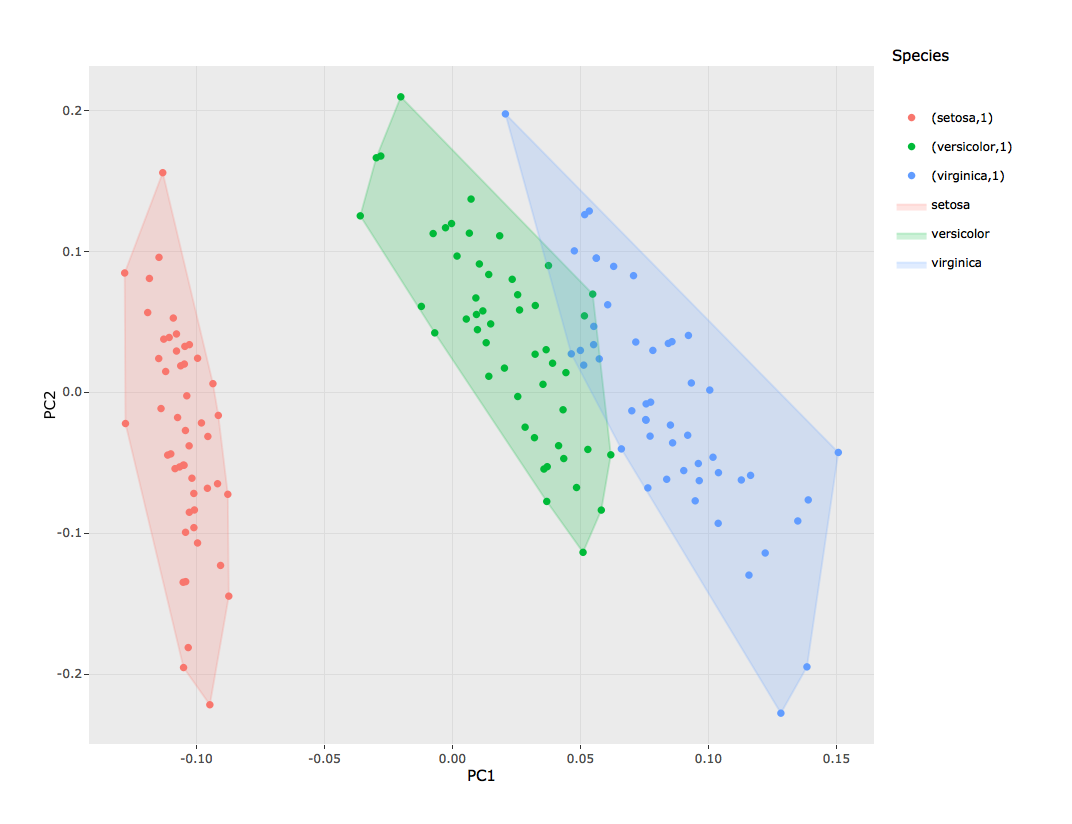}
  \caption{PCA with colors and boundary for each flower species.}
  \label{figure:iris_pca_full}
\end{figure}

The above example automatically plots the PCA results from \pkg{stats}
package. \code{autoplotly()} accepts parameters such as \code{frame} to
draw the boundaries for each flower species and \code{colour} to
indicate the column name to use to color each data points, as shown in
Figure \ref{figure:iris_pca_full}.

\begin{figure}[htbp]
  \centering
  \includegraphics[width=145mm,scale=0.8]{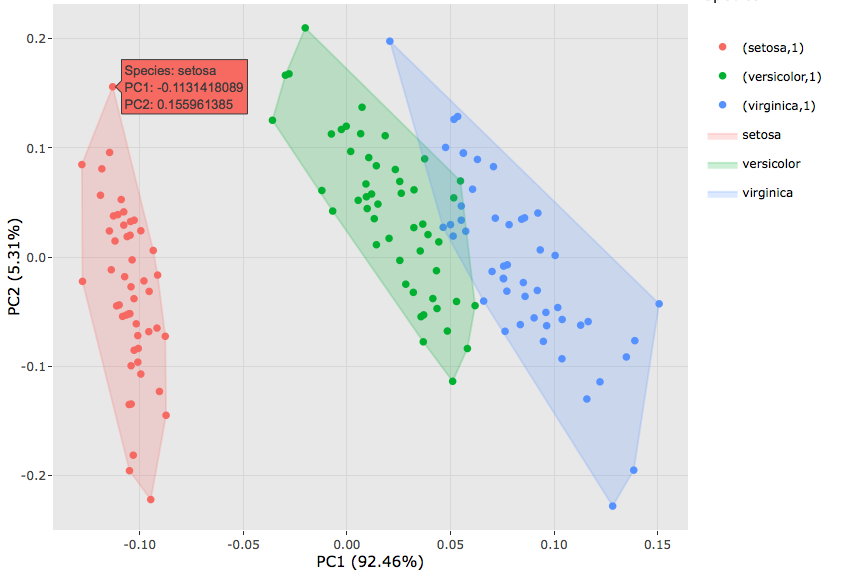}
  \caption{PCA with clolors and boundary for each principal component.}
  \label{figure:iris_pca_caption}
\end{figure}

Users can hover the mouse over to each point in the plot to see more
details, such as principal components information and the species this
particular data point belongs to, as shown in Figure
\ref{figure:iris_pca_caption}.

\begin{figure}[htbp]
  \centering
  \includegraphics[width=145mm,scale=0.8]{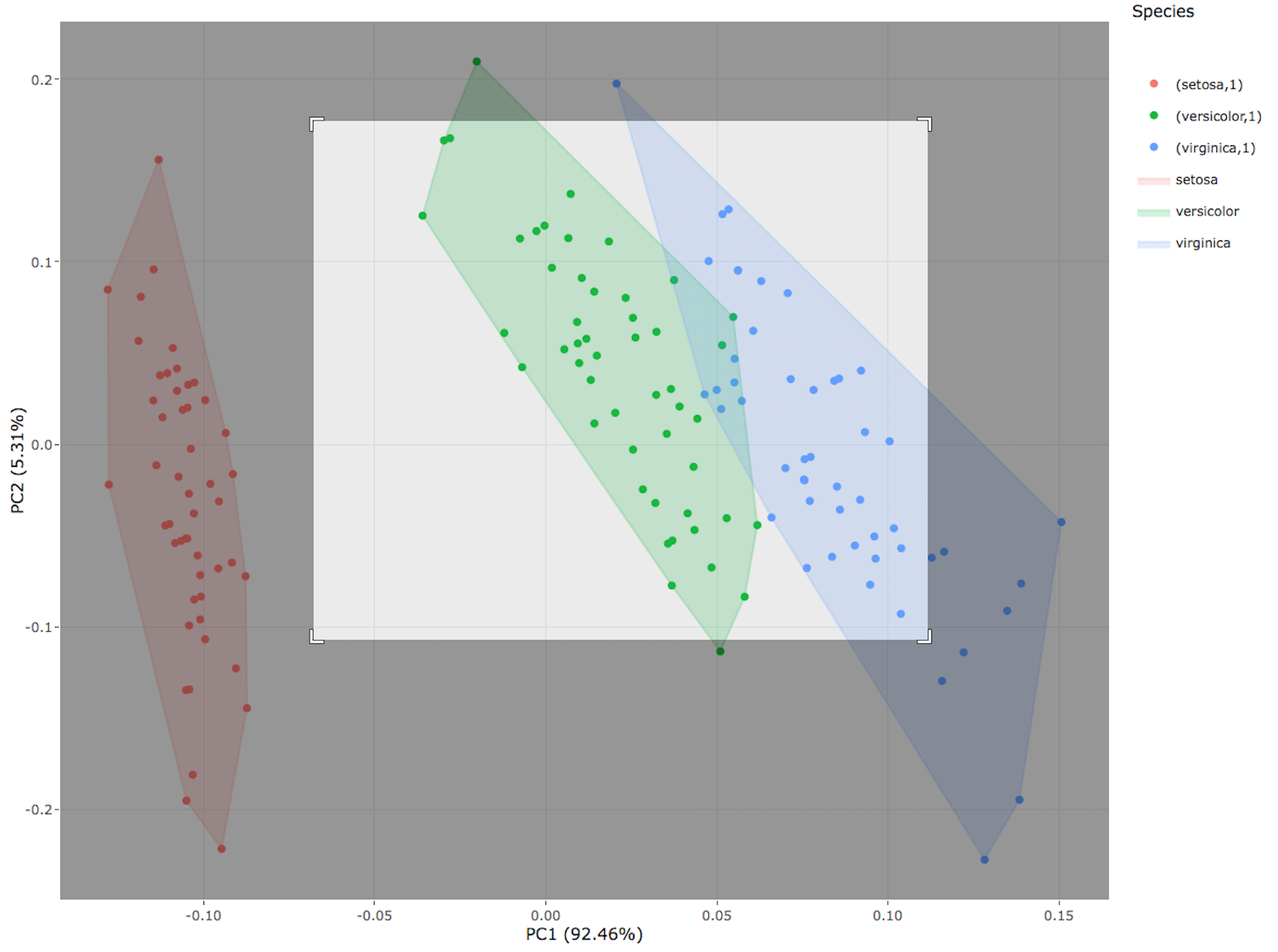}
  \caption{Zoom in using interactive selector.}
  \label{figure:iris_pca_zoom}
\end{figure}

Users can also use the interactive selector to drag and select an area
to zoom in, as shown in Figure \ref{figure:iris_pca_zoom}, and zoom out
by double clicking anywhere in the plot.

\subsection{Forecasting}

Forecasting packages such as \CRANpkg{forecast} \citep{forecast},
\CRANpkg{changepoint} \citep{changepoint}, \CRANpkg{strucchange}
\citep{strucchange}, and \CRANpkg{dlm} \citep{dlm}, are popular choices
for statisticians and researchers. Interactive visualizations of
predictions and statistical results from those packages can be generated
automatically using the functions provided by \pkg{autoplotly} with the
help of \pkg{ggfortify}.

\begin{figure}[htbp]
  \centering
  \includegraphics[width=145mm,scale=0.8]{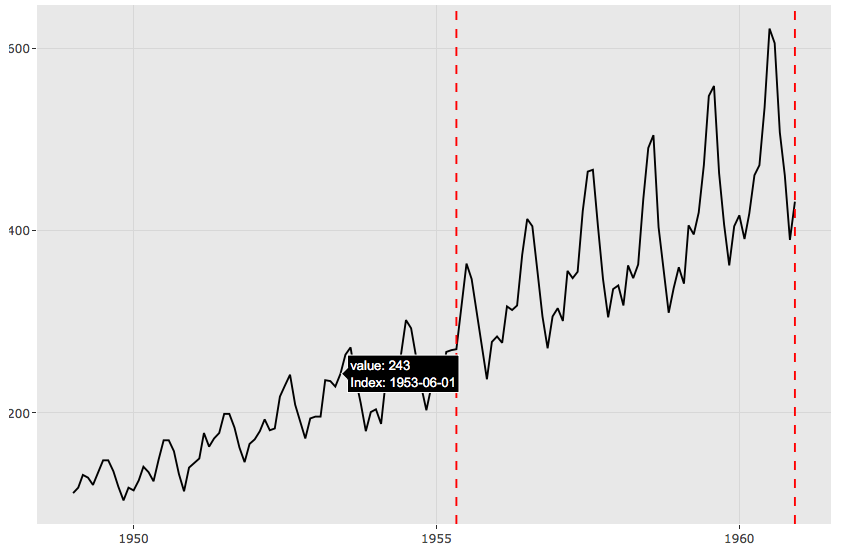}
  \caption{Change points with optimal positioning for AirPassengers.}
  \label{figure:changepoint_caption}
\end{figure}

The \pkg{autoplotly} function automatically plots the change points with
optimal positioning for the \code{AirPassengers} data set found in the
\pkg{changepoint} package using the \code{cpt.meanvar} function, shown
in Figure \ref{figure:changepoint_caption}.

\begin{Schunk}
\begin{Sinput}
library(changepoint)
autoplotly(cpt.meanvar(AirPassengers))
\end{Sinput}
\end{Schunk}

\begin{figure}[htbp]
  \centering
  \includegraphics[width=145mm,scale=0.8]{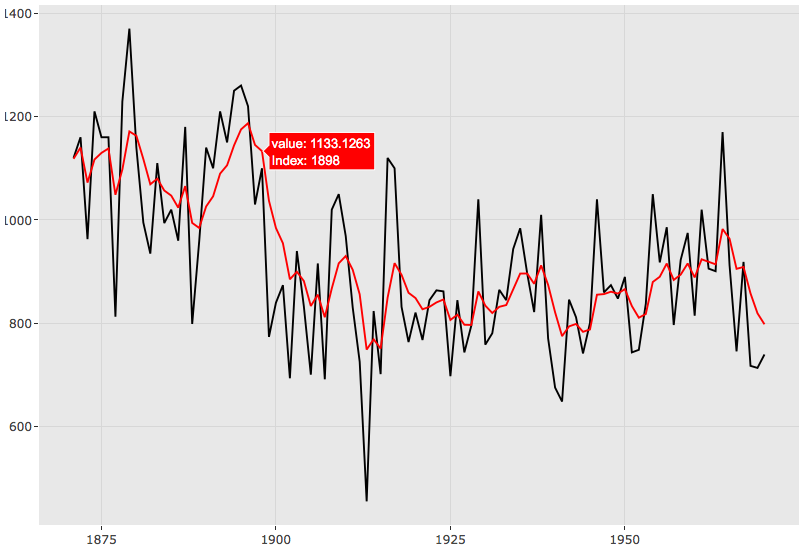}
  \caption{Smoothed time series by Kalman filter.}
  \label{figure:dlm_caption}
\end{figure}

The \pkg{autoplotly} function automatically plots the original and
smoothed line from Kalman filter function in \pkg{dlm} package as shown
in Figure \ref{figure:dlm_caption}.

\begin{Schunk}
\begin{Sinput}
library(dlm)
form <- function(theta){
  dlmModPoly(order = 1, dV = exp(theta[1]), dW = exp(theta[2]))
}
model <- form(dlmMLE(Nile, parm = c(1, 1), form)$par)
filtered <- dlmFilter(Nile, model)
autoplotly(filtered)
\end{Sinput}
\end{Schunk}

\begin{figure}[htbp]
  \centering
  \includegraphics[width=145mm,scale=0.8]{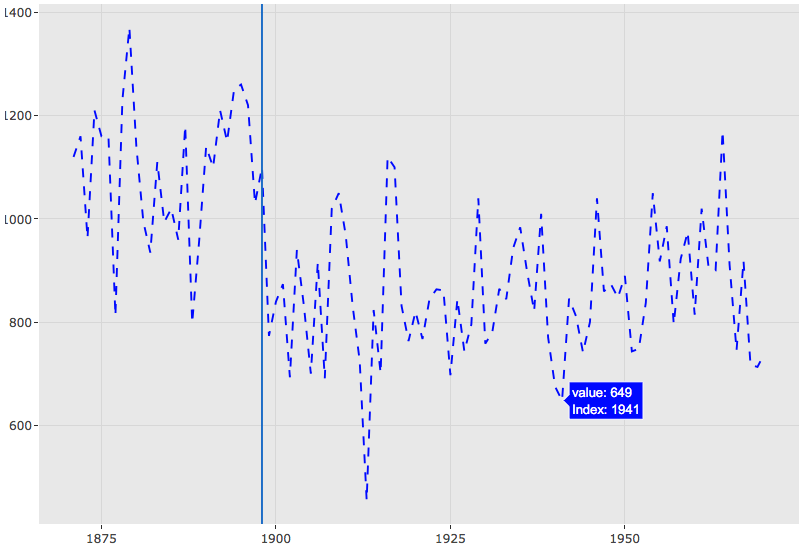}
  \caption{Optimal break points with possible structural changes.}
  \label{figure:strucchange_caption}
\end{figure}

Additionally, \pkg{autoplotly} plots the optimal break points where possible structural changes happen in the regression models built by the \code{strucchange::breakpoints}, shown in Figure \ref{figure:strucchange_caption}.

\begin{Schunk}
\begin{Sinput}
library(strucchange)
autoplotly(breakpoints(Nile ~ 1), ts.colour = "blue", ts.linetype = "dashed",
           cpt.colour = "dodgerblue3", cpt.linetype = "solid")
\end{Sinput}
\end{Schunk}

\subsection{Clustering}

\begin{figure}[htbp]
  \centering
  \includegraphics[width=145mm,scale=0.8]{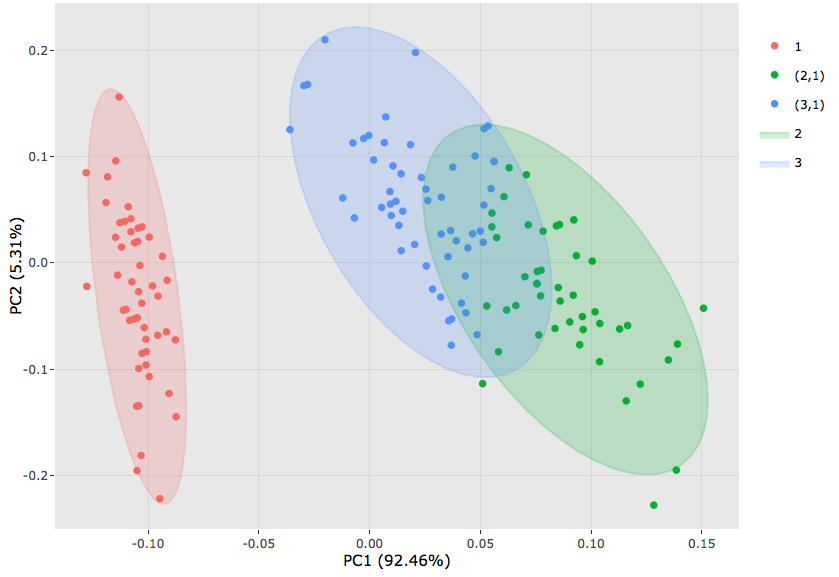}
  \caption{Fuzzy clustering of the data into 3 clusters.}
  \label{figure:cluster_caption}
\end{figure}

The \pkg{autoplotly} package also supports various objects like \code{cluster::clara}, \code{cluster::fanny}, \code{cluster::pam}, \code{stats::kmeans}, and \code{lfda::lfda}, from the \CRANpkg{cluster} \citep{cluster} and \CRANpkg{lfda} \citep{lfdaarxiv} packages. It automatically infers the object type and generate interactive plots of the results from those packages with a single function call. Users can specify \code{frame = TRUE} to easily draw the clustering boundaries as seen in Figure \ref{figure:cluster_caption} and the following code:

\begin{Schunk}
\begin{Sinput}
library(cluster)
autoplotly(fanny(iris[-5], 3), frame = TRUE, frame.type = "norm")
\end{Sinput}
\end{Schunk}

By specifying \code{frame.type}, users are
able to draw boundaries of different shapes. The different frame types can be found in \code{type} option in \code{ggplot2::stat\_ellipse}.

\subsection{Splines}

\begin{figure}[htbp]
  \centering
  \includegraphics[width=145mm,scale=0.8]{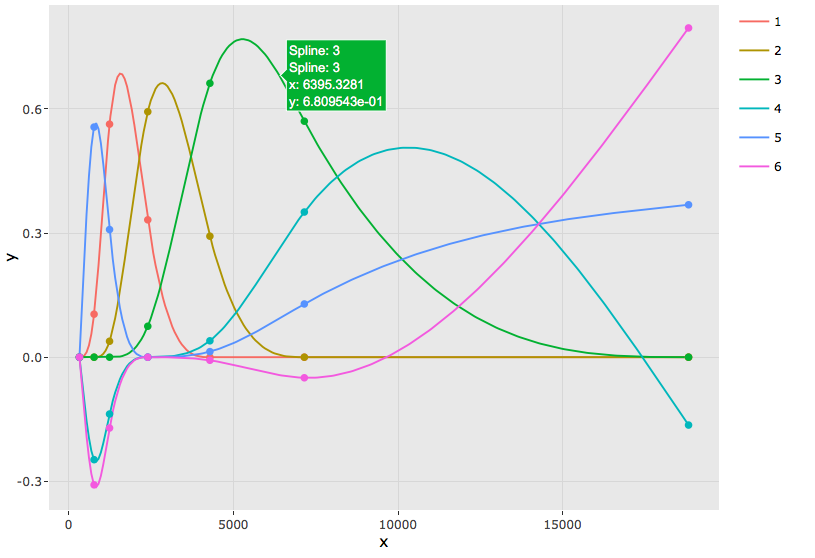}
  \caption{B-spline basis points for natural cubic spline with boundary knots.}
  \label{figure:splines_caption}
\end{figure}

The \code{autoplotly()} function can also automatically generate interactive plots for results producuced by \pkg{splines}, such as B-spline basis points visualization for natural cubic spline with boundary knots shown in Figure \ref{figure:splines_caption}.

\begin{Schunk}
\begin{Sinput}
library(splines)
autoplotly(ns(diamonds$price, df = 6))
\end{Sinput}
\end{Schunk}

\begin{figure}[htbp]
  \centering
  \includegraphics[width=145mm,scale=0.8]{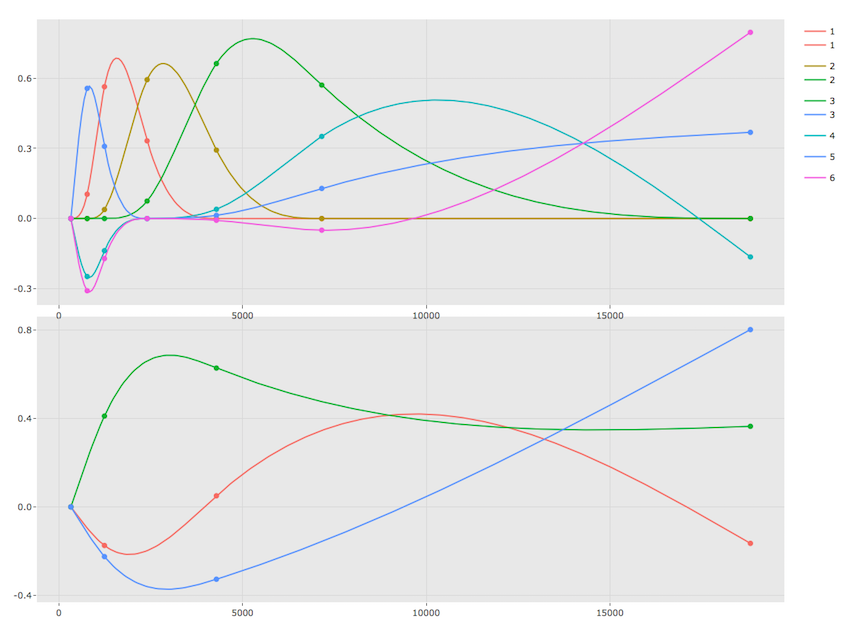}
  \caption{Multiple splines visualizations with different degree of freedom.}
  \label{figure:splines_subplot}
\end{figure}

Users can also stack multiple plots generated from \code{autoplotly()}
together in a single view using \code{subplot()}, two interactive
splines visualizations with different degree of freedom are stacked into
one single view in the following example, as shown in Figure
\ref{figure:splines_subplot}:

\begin{Schunk}
\begin{Sinput}
library(splines)
subplot(
  autoplotly(ns(diamonds$price, df = 6)),
  autoplotly(ns(diamonds$price, df = 3)), nrows = 2, margin = 0.01)
\end{Sinput}
\end{Schunk}

Note that the interactive control of those two plots are independent. In
other words, you can control two sub-plots separately without affecting
each other. Additional options for \code{subplot()} are available to
select whether to share axises, titles, and which layout of the plots to
adopt, etc.

\subsection{Linear models}

\begin{figure}[htbp]
  \centering
  \includegraphics[width=145mm,scale=0.8]{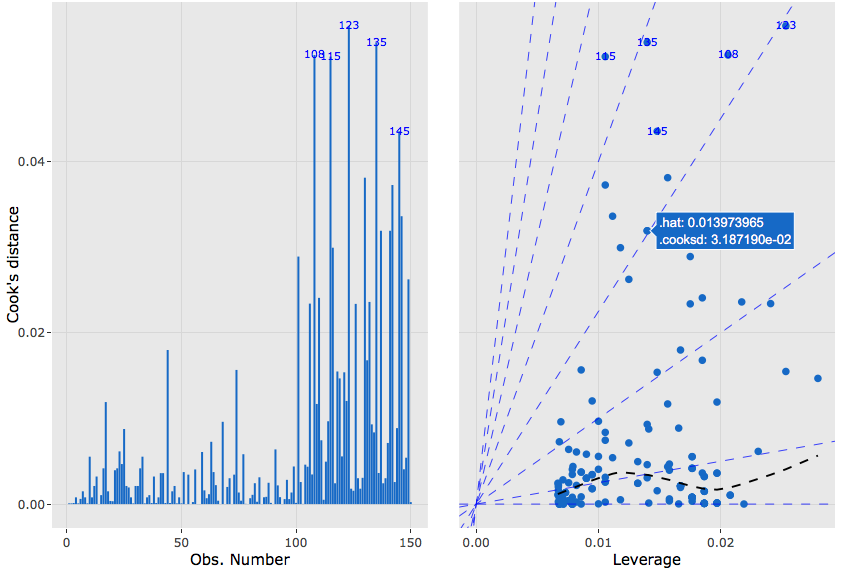}
  \caption{Linear model results.}
  \label{figure:lm_caption}
\end{figure}

The \code{autoplotly()} function is able able to interpret \code{lm}
fitted model objects and allows the user to select the subset of desired
plots through the \code{which} parameter (just like the \code{plot.lm}
function). The \code{which} parameter allows users to specify which of
the subplots to display. Many plot aesthetics can be changed by using
the appropriate named parameters. For example, the \code{colour}
parameter is for coloring data points, the \code{smooth.colour}
parameters is for coloring smoothing lines and the \code{ad.colour}
parameters is for coloring the auxiliary lines, as demonstrated in
Figure \ref{figure:lm_caption} and the following code:

\begin{Schunk}
\begin{Sinput}
autoplotly(
  lm(Petal.Width ~ Petal.Length, data = iris),
  which = c(4, 6), colour = "dodgerblue3",
  smooth.colour = "black", smooth.linetype = "dashed",
  ad.colour = "blue", label.size = 3, label.n = 5,
  label.colour = "blue")
\end{Sinput}
\end{Schunk}

\section{Future development}

\pkg{autoplotly} already supports a wide range of statistical results
from many popular packages. However, with recent advancements in
distributed systems and hardwares, many new frameworks have become more
and more popular to researchers with the help of those optimizations and
advancements, such as \CRANpkg{xgboost} \citep{xgboost} for scalable
gradient tree boosting, \CRANpkg{tensorflow} \citep{tensorflow},
\CRANpkg{tfestimators} \citep{tfestimatorspaper}, and \CRANpkg{keras}
\citep{keras} for machine learning that take advantages of hardware
accelerations. We'd like to support more types of statistical results
from more modern and scalable frameworks.

We welcome suggestions and contributions from others. With this package,
researchers will hopefully spend less time focusing on learning
\pkg{ggplot2} syntax or interactive visualization packages like
\pkg{plotly}. Instead they can spend more time on their work and
research. The source code of the package is located on Github
\url{https://github.com/terrytangyuan/autoplotly} where users can test
out development versions of the package and provide feature requests,
feedback and bug reports. We encourage you to submit your issues and
pull requests to help us make this package better for the R community.

\section{Summary}

The \pkg{autoplotly} package provides functionalities to automatically
generate interactive visualizations for many popular statistical results
supported by \pkg{ggfortify} package with \pkg{plotly} and \pkg{ggplot2}
style. The generated visualizations can also be easily extended using
\pkg{ggplot2} and \pkg{plotly} syntax while staying interactive. This
package allows users to spend more time on their research instead of
learning \pkg{ggplot2} syntax or interactive visualization packages like
\pkg{plotly}.

\bibliography{tang}

\address{%
Yuan Tang\\
H2O.ai\\
2309 Wake Robin Drive\\ West Lafayette, IN 47906\\
}
\href{mailto:terrytangyuan@gmail.com}{\nolinkurl{terrytangyuan@gmail.com}}

\end{article}

\end{document}